\documentclass[journal,draftclsnofoot,onecolumn,12pt,twoside]{IEEEtran}
\usepackage{times,amsmath,epsfig}
\usepackage{cite}
\usepackage{graphicx}
\usepackage{psfrag}
\usepackage{multicol}
\usepackage{float}
\usepackage{color}
\usepackage{url}
\usepackage{stfloats}
\usepackage{amsmath}
\usepackage{amssymb}
\usepackage{subcaption}
\usepackage{geometry}
\usepackage{caption}
\usepackage{amsthm}
\theoremstyle{remark}
\newtheorem{remark}{Remark}

\begin{document}

\title{X-duplex Relaying: Adaptive Antenna Configuration}
%\title{On the Performance of Full-Duplex Amplify-and-Forward Relay Systems}
%\author{Shuai~Li,
%        Mingxin~Zhou,
%        Jianjun~Wu,
%        Lingyang~Song,~\IEEEmembership{Senior~Member,~IEEE,}\\
%        Yonghui~Li,~\IEEEmembership{Senior~Member,~IEEE,}
%        and~Hongbin~Li}

\author{\IEEEauthorblockN{Shuai Li\IEEEauthorrefmark{1}, Mingxin Zhou\IEEEauthorrefmark{1}, Jianjun Wu\IEEEauthorrefmark{1}, Lingyang Song\IEEEauthorrefmark{1}, \\Yonghui Li\IEEEauthorrefmark{2}, and Hongbin Li\IEEEauthorrefmark{1}} \IEEEauthorblockA{\\\IEEEauthorrefmark{1} School of Electronics Engineering and Computer Science, \vspace{-2mm}\\Peking University, Beijing, China \vspace{-2mm}\\(E-mail:
{shuai.li.victor, mingxin.zhou, just, lingyang.song, lihb}@pku.edu.cn)\\ \IEEEauthorrefmark{2}School of Electrical and Information Engineering, \vspace{-2mm}\\The University of Sydney, Australia \vspace{-2mm}\\(E-mail: yonghui.li@sydney.edu.au)\\  }
}
\maketitle
 \thispagestyle{empty}
\pagestyle{empty}

\begin{abstract}
%\boldmath
In this letter, we propose a joint transmission mode and transmit/receive (Tx/Rx) antenna configuration scheme referred to as X-duplex in the relay network with one source, one amplify-and-forward (AF) relay and one destination.
The relay is equipped with two antennas, each of which is capable of reception and transmission.
In the proposed scheme, the relay adaptively selects its Tx and Rx antenna, operating in either full-duplex (FD) or half-duplex (HD) mode.
The proposed scheme is based on minimizing the symbol error rate (SER) of the relay system.
The asymptotic expressions of the cumulative distribution function (CDF) for the end-to-end signal to interference plus noise ratio (SINR), average SER and diversity order are derived and validated by simulations.
Results show that the X-duplex scheme achieves additional spatial diversity, significantly reduces the performance floor at high SNR and improves the system performance.
\end{abstract}

\vspace{-2mm}
\section{Introduction}

Deployment of full-duplex (FD) into relay networks is a promising technology to increase the spectral efficiency of wireless relay networks~\cite{Jic}.
The FD relays can receive and transmit the signal simultaneously over the same frequency which is contrary to the half-duplex (HD) relay systems requiring two orthogonal channels.
However, the performance of FD relaying is limited by self interference due to the signal leakage at the FD relay node.
Various approaches, including antenna isolation~\cite{Tr}, analog cancellation~\cite{Ee}, have been developed to mitigate the self interference and improve the performance of FD relay systems.

Adaptive mode selection between FD and HD is an effective way to further improve the system performance.
The hybrid FD/HD relaying has been investigated and shown that it can effectively improve spectral efficiency~\cite{Trhybird}.
The outage probability performance of an optimal relay selection scheme with hybrid relaying has been analyzed~\cite{Ik}.
A joint relay and antenna selection scheme (RAMS) has been proposed and considerably improved the system performance~\cite{Ky}.
The antenna switching has been introduced into the full-duplex wiretap channel to enhance physical layer security and obtain the full secrecy diversity order~\cite{sy}.
When the antennas at relay can be adaptively configured for transmission or reception, there are four possible transmission modes including two HD and two FD modes, as depicted in Fig. 1.
%However, the joint mode and antenna selection for FD relay system has not been well analyzed.
%In addition, the residual self interference (RSI) can be modeled as Rayleigh distribution in practical systems due to multi-path effect as ~\cite{Ik,Ky,HA}.
Due to the multi-path transmission of the signal leaked from the transmit antenna to the receive antenna at FD node, the residual self interference (RSI) can be modeled as Rayleigh distribution with effective self interference cancellation as~\cite{Ik,Ky,HA}.
In this case, the analysis becomes non-trivial.

In this letter, we consider a relay system which consists of one source, one destination and one amplify-and-forward (AF) relay.
Given antennas capable of transmission or reception, the relay can dynamically operate in four modes as shown in Fig. 1.
We propose to configure the antennas based on minimizing the symbol error rate (SER) of the system.
%the signal to interference plus noise ratio (SINR)
%The relay is equipped with two shared-antennas, which are able to transmit and receive~\cite{Db}.
%We propose the X-duplex scheme, where the system adaptively selects the optimal antenna to receive/transmit signal, and switches between FD and HD mode to achieve the maximum average sum rate based on the instantaneous channel state information (CSI) and residual self interference.
The asymptotic cumulative distribution function (CDF) of the end-to-end signal to interference plus noise ratio (SINR) at the destination is derived.
Based on the CDF, the asymptotic average SER and diversity order are derived and validated by numerical simulations.
Results show that the X-duplex scheme achieves additional spatial diversity, significantly reduces the performance floor at high SNR and improves the system performance.

\IEEEpeerreviewmaketitle

\vspace{-2mm}
\section{System Model}

\begin{figure}
        \begin{subfigure}[b]{0.475\textwidth}
            \includegraphics[width=\textwidth]{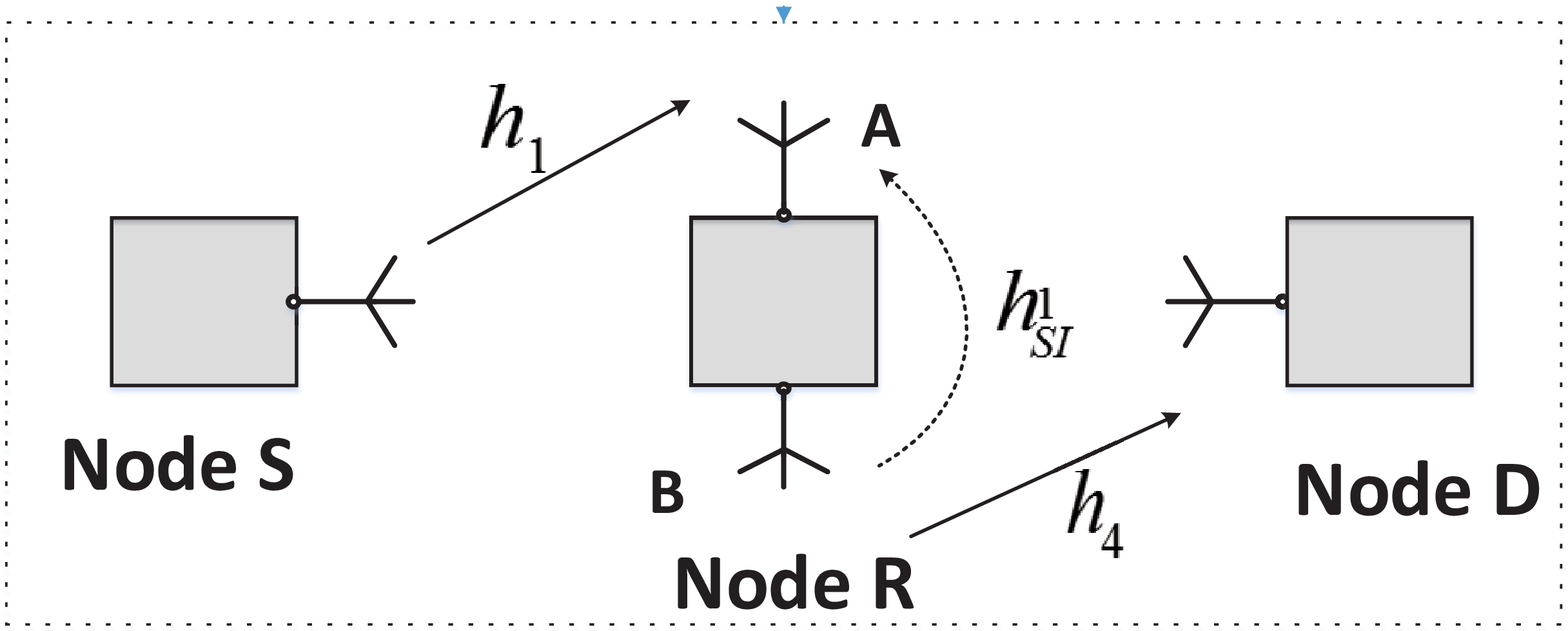}
            \caption[FD mode A]%
            {{\small FD mode A}}
            \label{fig:FD mode A}
        \end{subfigure}
        \quad
        \begin{subfigure}[b]{0.475\textwidth}
            \includegraphics[width=\textwidth]{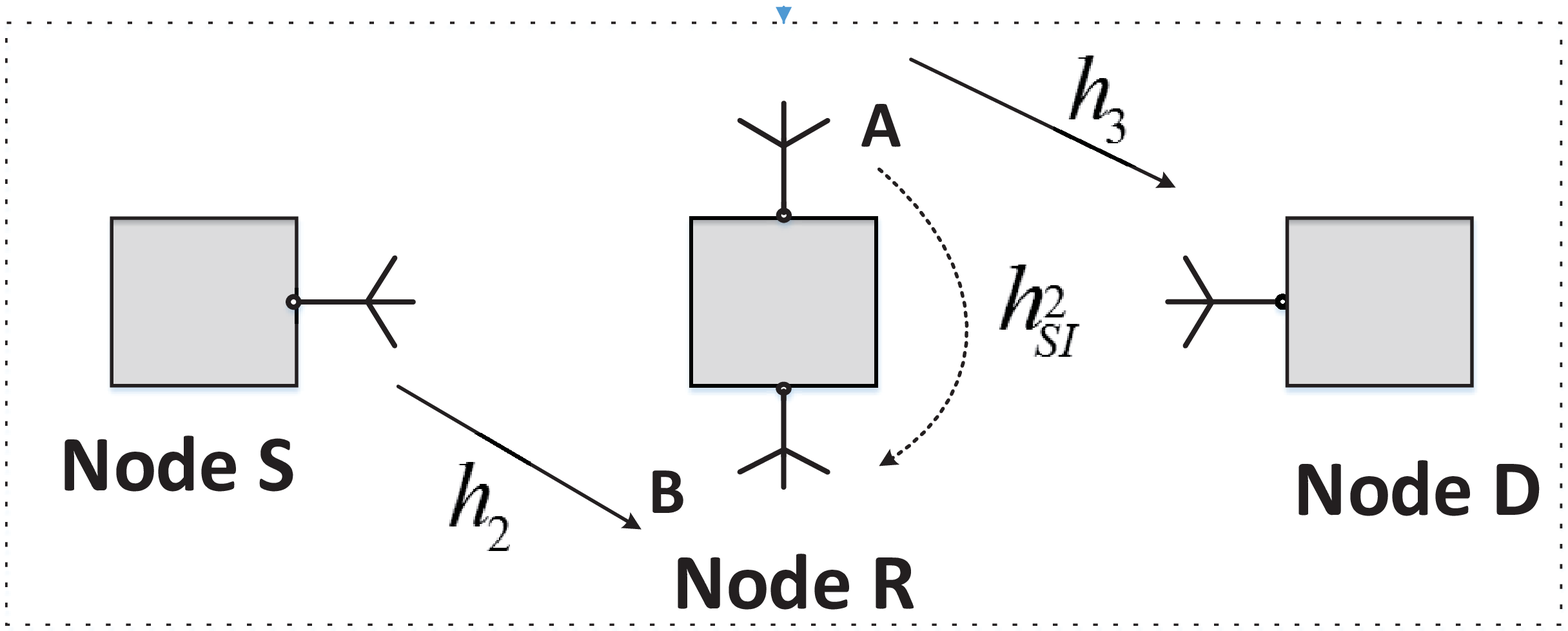}
            \caption[FD mode B]%
            {{\small{FD mode B}}}
            \label{fig:FD mode B}
        \end{subfigure}
        \vskip\baselineskip
        \vspace{-4mm}
        \begin{subfigure}[b]{0.475\textwidth}
            \includegraphics[width=\textwidth]{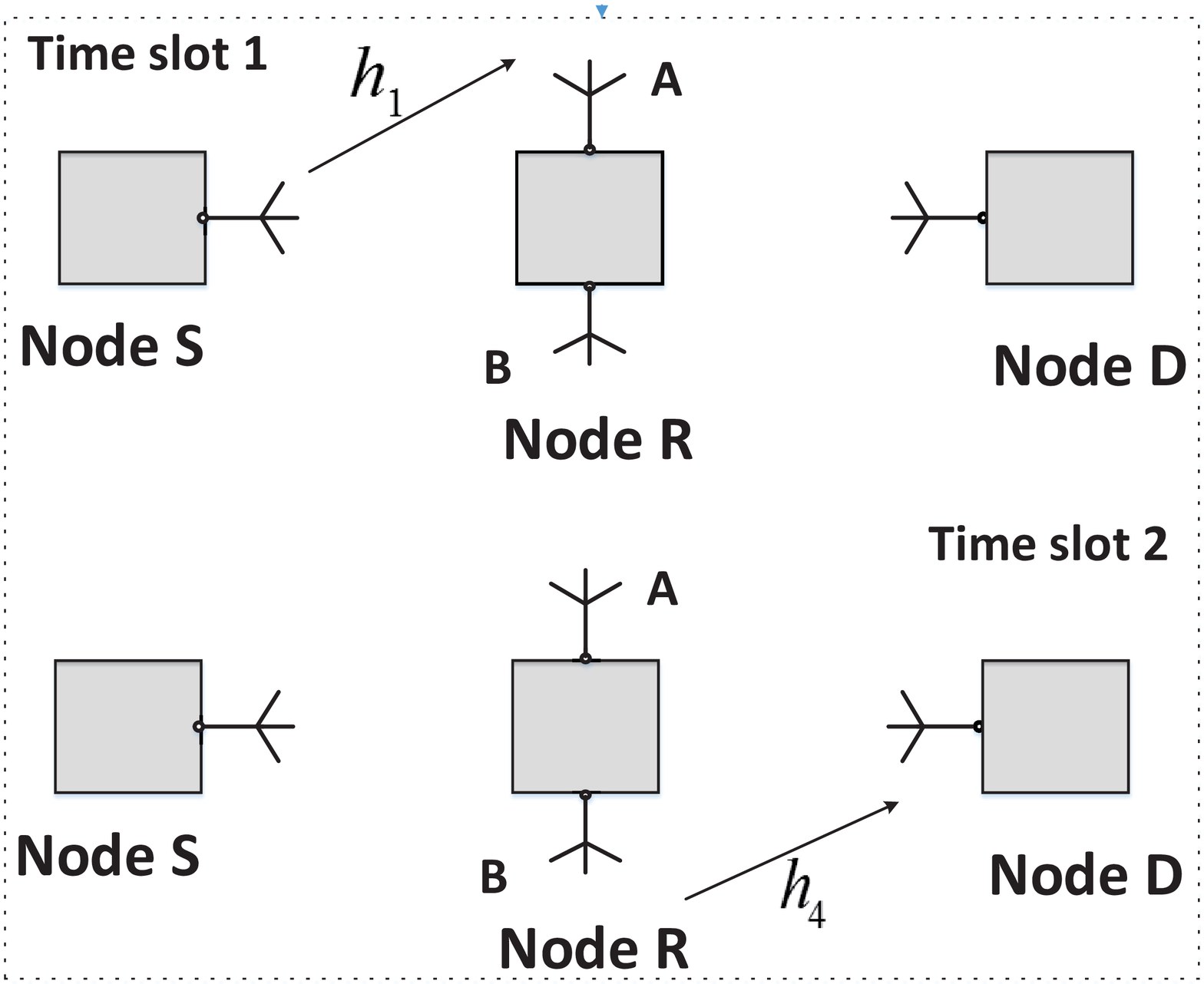}
            \caption[HD mode A]%
            {{\small HD mode A}}
            \label{fig:HD mode A}
        \end{subfigure}
        \quad
        \begin{subfigure}[b]{0.475\textwidth}
            \includegraphics[width=\textwidth]{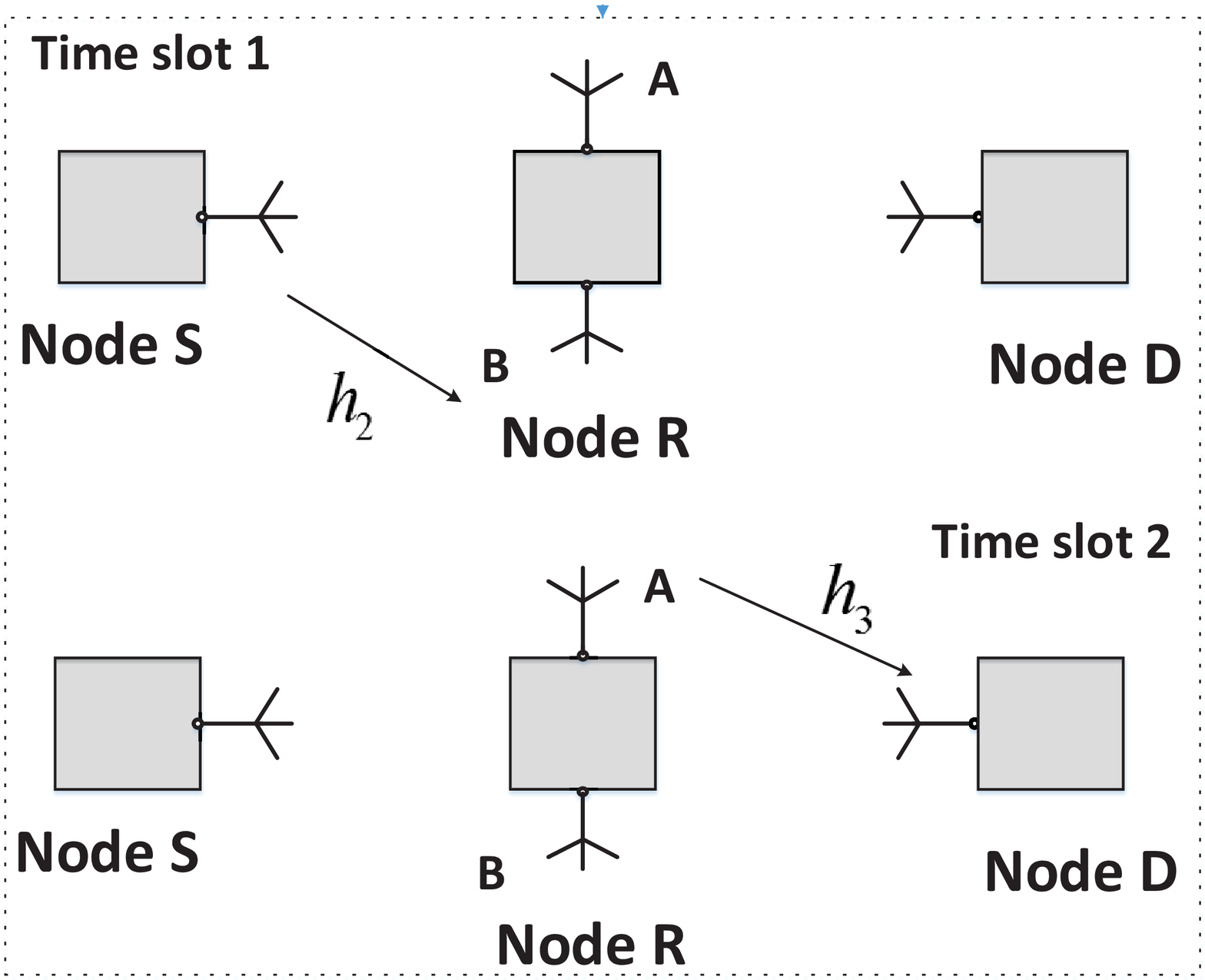}
            \caption[HD mode B]%
            {{\small HD mode B}}
            \label{fig:HD mode B}
        \end{subfigure}
        \caption[ System model ]
        {\small System model}
        \label{fig:System model}
        \vspace{-5mm}
    \end{figure}

%In this paper, we assume a RF chain number preserved condition.
%There are two RF chains in the relay node, one TX chain for transmitting signal and one RX chain for  receiving signal, respectively.
%The RF chains can be adaptively configured to connect with the antennas by the relay node.
%Based on the instantaneous channel state information (CSI) and residual self interference (RSI), the relay chooses which antenna to transmit or receive and configures the each RF chain to connect with the assigned antenna in the beginning of every time slot.
%Meanwhile, the relay chooses which duplex mode to operate.
%In this way, the optimal operating mode is selected from the four modes, i.e., FD mode A, FD mode B, HD mode A, HD mode B.
%Also, we assume the assigned transmit and receive antenna remain unchanged in one time slot, to avoid the influence of RF chain halfway switch on the quality of the signal and reduction of energy efficiency.
In this paper, we consider a two-hop relay system with one source node (S), one AF relay node (R) and one destination node (D), as shown in Fig. \ref{fig:System model}.
The source S transmits the information to the destination D with the help of the relay R, and all nodes operate at the same frequency.
The relay R is equipped with two antennas, denoted by A and B, where each antenna is able to transmit/receive the signal.
Based on the instantaneous channel state information (CSI) and RSI, the relay adaptively chooses which antenna to transmit/receive to optimize the system performance.
For simplicity and without loss of generality, we assume that Tx and Rx antenna at relay remain unchanged in one time slot.

This system can operate in four modes, FD mode A, HD mode A, FD mode B and HD mode B.
In FD mode A, relay R set antenna A as Rx antenna and antenna B as Tx antenna.
Relay R operates at FD mode, i.e., antenna A receives and antenna B transmits signal simultaneously in one time slot.
In HD mode A, the Tx/Rx antenna is the same and relay R operates at HD mode: source S transmits the signal to relay in the first half of one time slot and relay forwards the signal to destination D in the second half of one time slot.
In FD/HD mode B, the Tx/Rx antenna at relay is swapped compared with FD/HD mode A.%The FD mode B and HD mode B are also shown in Fig.1.
%\textbf{FD mode B}: relay R set antenna B as receiving antenna and antenna A as transmitting antenna, and operates at FD mode;
%\textbf{HD mode B}: relay R set antenna B as receiving antenna and antenna A as transmitting antenna and operates at HD mode.

The channels between the source and relay are denoted as $h_1, h_2$ and the channels between the relay and destination are denoted as $h_3, h_4$.
The RSI channel from antenna B to antenna A is denoted as $h_{SI}^1$, and the RSI channel from antenna A to antenna B as $h_{SI}^2$.
All the channels are assumed to follow block Rayleigh fading, where each channel remains unchanged in one time slot and varies from one slot to another independently~\cite{Ik,Ky,HA}.
The SINR of each mode can be expressed as
\begin{small}\begin{eqnarray}
{\gamma _{{fd_a}}} &=& \frac{{{X_1}{P_R}{\gamma _4}}}{{{X_1} + {P_R}{\gamma _4} + 1}},
{\gamma _{{hd_a}}} = \frac{{{P_S}{P_R}{\gamma _1}{\gamma _4}}}{{{P_S}{\gamma _1} + {P_R}{\gamma _4} + 1}},\nonumber\\
{\gamma _{fd_b}} &=& \frac{{{X_2}{P_R}{\gamma _3}}}{{{X_2} + {P_R}{\gamma _3} + 1}},
{\gamma _{hd_b}} = \frac{{{P_S}{P_R}{\gamma _2}{\gamma _3}}}{{{P_S}{\gamma _2} + {P_R}{\gamma _3} + 1}},
\end{eqnarray}\end{small}
\hspace{-1mm}where $P_S$ and $P_R$ are the transmit power of the source and relay.
$n_R$ is the additive white Gaussian noise (AWGN) with variance ${\sigma ^2}$.
${\gamma _i} = {\left| {{h_i}} \right|^2}/{\sigma ^2},i \in \{1,2,3,4 \}$, ${\gamma ^j}_{SI} = |h_{SI}^j{|^2}/{\sigma ^2}, j \in \{1,2 \}$, ${X_1} = \frac{{{P_S}{\gamma _1}}}{{{P_R}{\gamma ^1}_{SI} + 1}}$, ${X_2} = \frac{{{P_S}{\gamma _2}}}{{{P_R}{\gamma ^2}_{SI} + 1}}$.
The channel gain ${\gamma _i}$ is modeled as exponential distribution with average mean $\lambda_i$
and SNR ${\gamma ^j}_{SI}$ follows Rayleigh distribution with average mean ${\lambda_R^j}$.
%In this paper, we assume the two RSI channels independent and identically distributed (i.i.d) due to the same self interference cancellation method with the same configurations
In this paper, we assume the two RSI channels identical due to the same self interference cancellation module in two FD modes, ${\lambda_R^1}$ = ${\lambda_R^2}$.
The CSIs of $h_i$ and $h_{SI}^j$ can be measured by standard pilot-based channel estimation and sufficient training, and transmitted to the decision node through reliable feedback channels~\cite{Trhybird,Ik}.
We assume perfect CSI in this paper.

%As the average sum rate of the HD mode can be given as
%${R_i} = \frac{1}{2}{\log _2}(1 + {\gamma _i}),i = h{d_a},h{d_b}$, the equivalent end-to-end SINR of HD mode can be written as $\sqrt {{\gamma_i } + 1}  - 1,i = h{d_a},h{d_b}$.
%%\subsection{X-duplex Relay}

To optimize the system performance, the X-Duplex can be reduced to one of the four modes with different antenna mode configurations.
The average SER with link SINR $\gamma$ can be written as $\overline {SER} = {a_1}{\mathbb{E}}[Q(\sqrt {2{a_2}\gamma } )]$ \cite{AJ}.
The optimal Tx/Rx mode and duplex mode is determined based on the minimal SER criterion.
\begin{small}\begin{equation}\label{mode}
  mode = \arg \min \left\{ {\overline {SER} } \right\} = \arg \max \left\{ {{{\tilde \gamma }_{f{d_a}}},{{\tilde \gamma }_{f{d_b}}},{{\tilde \gamma }_{h{d_a}}},{{\tilde \gamma }_{h{d_b}}}} \right\},
\end{equation}\end{small}
\hspace{-1mm}where $\tilde \gamma$ is the end-to-end SINR of each  mode.
As the equivalent end-to-end SINR of HD mode can be written as $\tilde \gamma  = \sqrt {\gamma  + 1}  - 1$~\cite{Trhybird}, the SINR of the X-Duplex relay system can be given as
\begin{small}\begin{equation}
{\gamma _{\max }} = \max \{ {\gamma _{fd_a}},{\gamma _{fd_b}},\sqrt {{\gamma _{hd_a}} + 1}  - 1,\sqrt {{\gamma _{hd_b}} + 1}  - 1\} .
\end{equation}\end{small}

\vspace{-5mm}
\section{Performance Analysis}

In this section, we derive the CDF of the X-duplex relay system and analyze the system performance, including the average SER and diversity order.

\vspace{-6mm}
\subsection{Average SER Analysis}

The average SER with link SINR $\gamma$ can be computed as \cite{AJ}
\begin{small}\begin{equation}\label{ser}
\overline {SER}  = {a_1}{\mathbb{E}}[Q(\sqrt {2{a_2}\gamma } )] = \frac{{{a_1}\sqrt {{a_2}} }}{{2\sqrt \pi  }}\int\limits_0^\infty  {\frac{{{e^{ - {a_2}\gamma }}}}{{\sqrt \gamma  }}{F_\gamma }(\gamma )d\gamma },
\end{equation}\end{small}
where ${F_\gamma }( \cdot )$ is the CDF of $\gamma $, $Q( \cdot )$ is the Gaussian Q-Function \cite{zwillinger2014table}, $({a_1},{a_2})$ denote the modulation formats.

\emph{Proposition 1:} The asymptotic CDF of the end-to-end SINR of the X-duplex system $\gamma_{\max}$ can be given as
\begin{small}\begin{equation}\label{eq:cdf}
 \Pr ({\gamma _{\max }} < x) = \left( {1 - {I_1} + {I_2}} \right) \cdot \left( {1 - {I_3} + {I_4}} \right),
\end{equation}
\begin{eqnarray}
  &&{I_1} = \frac{1}{{1 + {\eta _1}x}}\left[ {\beta _1^1{K_1}(\beta _1^1){e^{ - {C_1}x}} + {\eta _1}x\beta _2^1{K_1}(\beta _2^1){e^{ - \beta _3^1}}} \right],\nonumber\\
  &&{I_2} = \frac{{2{\eta _1}({x^2} + x)}}{{{\lambda _4}{P_R}{{(1 + {\eta _1}x)}^2}}}\left[ {{K_0}(\beta _1^1){e^{ - {C_1}x}} - {K_0}(\beta _2^1){e^{ - \beta _3^1}}} \right],\nonumber\\
  &&{I_3} = \frac{1}{{1 + {\eta _2}x}}\left[ {\beta _1^2{K_1}(\beta _1^2){e^{ - {C_2}x}} + {\eta _2}x\beta _2^2{K_1}(\beta _2^2){e^{ - \beta _3^2}}} \right],\nonumber\\
  &&{I_4} = \frac{{2{\eta _2}({x^2} + x)}}{{{\lambda _3}{P_R}{{(1 + {\eta _2}x)}^2}}}\left[ {{K_0}(\beta _1^2){e^{ - {C_2}x}} - {K_0}(\beta _2^2){e^{ - \beta _3^2}}} \right],\nonumber
\end{eqnarray}
\vspace{-4mm}
\begin{align}
&\beta _1^1 = 2\sqrt {\frac{{x + {x^2}}}{{{\lambda _1}{\lambda _4}{P_S}{P_R}}}}, \beta _3^1 = {C_1}({x^2} + 2x) + \frac{{x + 1}}{{{\eta _1}{\lambda _1}{P_S}}}, \nonumber\\
&\beta _1^2 = 2\sqrt {\frac{{x + {x^2}}}{{{\lambda _2}{\lambda _3}{P_S}{P_R}}}}, \beta _3^2 = {C_2}({x^2} + 2x) + \frac{{x + 1}}{{{\eta _2}{\lambda _2}{P_S}}},\nonumber
\end{align}\end{small}
\vspace{-5mm}
\begin{small}\begin{align}
&\beta _2^1 = 2\sqrt {\frac{{{{({x^2} + 2x)}^2} + {x^2} + 2x + \frac{1}{{{\eta _1}}}(x + 1)({x^2} + 2x)}}{{{\lambda _1}{\lambda _4}{P_S}{P_R}}}},\nonumber\\
&\beta _2^2 = 2\sqrt {\frac{{{{({x^2} + 2x)}^2} + {x^2} + 2x + \frac{1}{{{\eta _2}}}(x + 1)({x^2} + 2x)}}{{{\lambda _2}{\lambda _3}{P_S}{P_R}}}},\nonumber
\end{align}\end{small}
where ${\eta _1} = \frac{{\lambda _R^1{P_R}}}{{{\lambda _1}{P_S}}}$, ${C_1} = \frac{1}{{{\lambda _1}{P_S}}} + \frac{1}{{{\lambda _4}{P_R}}}$, ${\eta _2} = \frac{{\lambda _R^2{P_R}}}{{{\lambda _2}{P_S}}}$, ${C_2} = \frac{1}{{{\lambda _2}{P_S}}} + \frac{1}{{{\lambda _3}{P_R}}}$, ${K_1}( \cdot )$, ${K_0}( \cdot )$ are the first and zero order Bessel function of the first and second kind~\cite{zwillinger2014table}.

\begin{IEEEproof}
The deviation is given in Appendix A.
\end{IEEEproof}

\emph{Proposition 2:} The asymptotic average SER of the X-duplex system $\gamma_{\max}$ is given in \eqref{eq:ser}.

In \eqref{eq:ser}, ${\mu _1} = 2{C_1} + {a_2} + \frac{1}{{{\lambda _1}{P_S}{\eta _1}}}$, $\mu _1^2 = 2{C_2} + {a_2} + \frac{1}{{{\lambda _2}{P_S}{\eta _2}}}$, ${\mu _2} = {\mu _1} + {C_2}$,$\mu _2^2 = \mu _1^2 + {C_1}$,${\mu _3} = 2{C_1} + 2{C_2} + {a_2} + \frac{1}{{{\lambda _1}{P_S}{\eta _1}}} + \frac{1}{{{\lambda _2}{P_S}{\eta _2}}}$,
${\mathcal F^2}(v,\beta ,\gamma ) = {\eta _1}\mathcal F(v,\beta ,\gamma ,{\eta _1}) - {\eta _2}\mathcal F(v,\beta ,\gamma ,{\eta _2})$,
${\mathcal F^3} = \sqrt {{\eta _1}} {e^{\frac{{{C_1} + {C_2}}}{{{\eta _1}}}}}\Gamma (\frac{1}{2},\frac{{{C_1} + {C_2}}}{{{\eta _1}}}) - \sqrt {{\eta _2}} {e^{\frac{{{C_1} + {C_2}}}{{{\eta _2}}}}}\Gamma (\frac{1}{2},\frac{{{C_1} + {C_2}}}{{{\eta _2}}})$,
$\Gamma (\cdot)$ is the Gamma Function, $\Gamma (a,x)$ is the incomplete Gamma Function, ${D_{p}}( \cdot )$ is the Parabolic Cylinder Function \cite{zwillinger2014table}.

\begin{IEEEproof}
The deviation is given in Appendix B.
\end{IEEEproof}

\begin{figure*}[!t]
\vspace{-2mm}
\begin{small}
\setcounter{equation}{5}
\begin{align}\label{eq:ser}
&\overline {SER} \approx \frac{{{a_1}\sqrt {{a_2}} }}{{2\sqrt \pi  }}\left\{ {\sqrt {\frac{\pi }{{{a_2}}}}  - \sqrt {\frac{\pi }{{{\eta _1}}}} {e^{\frac{{{a_2} + {C_1}}}{{{\eta _1}}}}}\Gamma (\frac{1}{2},\frac{{{a_2} + {C_1}}}{{{\eta _1}}}) - \sqrt {\frac{\pi }{{{\eta _2}}}} {e^{\frac{{{a_2} + {C_2}}}{{{\eta _2}}}}}\Gamma (\frac{1}{2},\frac{{{a_2} + {C_2}}}{{{\eta _2}}}) + \frac{{\sqrt \pi  }}{{{\eta _1} - {\eta _2}}}{\mathcal F^3}- {\eta _1}{e^{ - \frac{1}{{{\lambda _1}{P_S}{\eta _1}}}}}\mathcal F(\frac{3}{2},{C_1},{\mu _1},{\eta _1})   } \right.
\nonumber\\
& \left. { { - {\eta _2}{e^{ - \frac{1}{{{\lambda _2}{P_S}{\eta _2}}}}}\mathcal F(\frac{3}{2},{C_2},\mu _1^2,{\eta _2})} +\hspace{-1mm} \frac{{{\eta _1}{e^{ - \frac{1}{{{\lambda _1}{P_S}{\eta _1}}}}}}}{{{\eta _1} - {\eta _2}}}{\mathcal F^2}(\frac{3}{2},{C_1},{\mu _2}) +\hspace{-1mm} \frac{{{\eta _2}{e^{ - \frac{1}{{{\lambda _2}{P_S}{\eta _2}}}}}}}{{{\eta _1} - {\eta _2}}}{\mathcal F^2}(\frac{3}{2},{C_2},\mu _2^2) +\hspace{-1mm} \frac{{{\eta _1}{\eta _2}{e^{ - \frac{1}{{{\lambda _1}{P_S}{\eta _1}}} -\hspace{-1mm} \frac{1}{{{\lambda _2}{P_S}{\eta _2}}}}}}}{{{\eta _1} - {\eta _2}}}{\mathcal F^2}(\frac{5}{2},{C_2},{\mu _3})} \right\},\\
&\mathcal F(v,\beta ,\gamma ,t) = {\left( {2\beta } \right)^{ - \frac{v}{2}}}\Gamma \left( v \right){D_{ - v}}(\frac{{\gamma  + t}}{{\sqrt {2\beta } }}){e^{\frac{{{{(\gamma  + t)}^2}}}{{8\beta }}}} - \frac{1}{2}{t^2}{\left( {2\beta } \right)^{ - \frac{v}{2} - 1}}\Gamma (v + 2){D_{ - v - 2}}(\frac{{\gamma  + \frac{5}{3}t}}{{\sqrt {2\beta } }}){e^{\frac{{{{(\gamma  + \frac{5}{3}t)}^2}}}{{8\beta }}}},
\end{align}
\setcounter{equation}{7}
\hrulefill
\vspace*{4pt}
\end{small}
\vspace{-6mm}
\end{figure*}
When SNR goes infinite, the CDF of FD mode A approaches $\Pr (\gamma _{fd_a} < x) = 1 - \frac{1}{{1 + \eta_1 x}}$. With (\ref{ser}) and \cite[eq.(3.383.10)]{zwillinger2014table}, the lower bound of the SER can be obtained as ${\overline {SER} _{SNR \to \infty }}= \frac{{{a_1}\sqrt {{a_2}} }}{{2\sqrt {\pi {\eta _1}} }}{e^{\frac{1}{{{\eta _1}}}{a_2}}}\Gamma (\frac{3}{2})\Gamma ( - \frac{1}{2},\frac{1}{\eta_1 }{a_2})$.
Compared with FD mode, the X-duplex scheme reduces the error floor and achieves lower SER in the high SNR region.

\subsection{Diversity Order Analysis}

According to \cite[eq.(10.30)]{Frank}, when $z$ comes close to zero, ${K_1}(z)$ function converges to $\frac{1}{z}$, and the value of ${K_0}(z)$ is comparatively small. Therefore,  at high SNR, the outage probability of X-duplex relay system can be approximated as
\begin{small}\begin{equation}\label{pout}
{P_{out}}(x) =  ( {1 -\hspace{-1mm} \frac{{{e^{ - {C_1}x}} + {\eta _1}x{e^{ - \beta _3^1}}}}{{1 + {\eta _1}x}}} ) \cdot ( {1 -\hspace{-1mm}  \frac{{{e^{ - {C_2}x}} + {\eta _2}x{e^{ - \beta _3^2}}}}{{1 + {\eta _2}x}}} ),
\end{equation}\end{small}
\hspace{-1.5mm}when SNR goes infinite, the outage probability of X-duplex relay system comes to zero.

We assume the identical transmit power of source and relay, ${P_S} = {P_R} = P_t$ and $\lambda_1=\lambda_2, \lambda_3 =\lambda_4, \lambda_R^1 =\lambda_R^2$, the finite SNR diversity order of X-duplex system can be derived with $
d(\lambda ) =  - \frac{{\partial \ln {P_{out}}(\lambda )}}{{\partial \ln \lambda }} =  - \frac{\lambda }{{{P_{out}}(\lambda )}}\frac{{\partial {P_{out}}(\lambda )}}{{\partial \lambda }}$ \cite{Ky} as
\begin{small}\begin{equation}\label{do}
{d_{XD}} \approx\hspace{-1mm} \frac{{P_t} \cdot {\partial [2(1 + \eta_1 x)M - {M^2}]/\partial {P_t}}}{{{{(1 + \eta_1 x)}^2}\hspace{-1mm} - 2(1 + \eta_1 x)M + {M^2}}},M = {e^{ - \frac{{2{\rho _1}}}{{{P_t}}}}} + \eta_1 x{e^{ - \frac{{2{\rho _2}}}{{{P_t}}}}},\end{equation}\end{small}
\hspace{-2mm}where ${C_3} = \frac{1}{{{\lambda _1}}} + \frac{1}{{{\lambda _4}}}$, ${\rho _1} = {C_3}x$, ${\rho _2} = {C_3}({x^2} + 2x) + \frac{{x + 1}}{{{\eta _1}{\lambda _1}}}$.
At high SNR, with Taylor's formula ${e^{ - x}} \approx 1 - x + \frac{1}{2}{x^2} - \frac{1}{6}{x^3} +...$, we can derive $d_{XD} \approx \frac{1}{{{P_t}}}\frac{{\frac{1}{{{P_t}}}(2{\rho _1}^2 + 2{\eta _1}^2{x^2}{\rho _2}^2{\rm{ + }}4{\eta _1}x{\rho _1}{\rho _2}) + o({P_t}^{ - 2})}}{{\frac{1}{{{P_t}^2}}({\rho _1}^2 + {\eta _1}^2{x^2}{\rho _2}^2{\rm{ + }}2{\eta _1}x{\rho _1}{\rho _2}) + o({P_t}^{ - 3})}}$.
When SNR goes infinite, $d_{XD}$ approaches two.

With \eqref{eq:hda}, \eqref{eq:fda}, \eqref{eq:hs} and Taylor's formula ${e^{ - x}} \approx 1 - x$, the diversity order of HD mode A, FD mode A and hybrid FD/HD scheme (HY) proposed in~\cite{Trhybird} can be derived as
\begin{small}\begin{align}
&{d_{H{D_a}}} = \frac{1}{{{P_t}}}\frac{{{C_3}({x^2} + 2x) \cdot {e^{ - {C_1}({x^2} + 2x)}}}}{{1 - {e^{ - {C_1}({x^2} + 2x)}}}} \approx 1 - \frac{{{C_3}({x^2} + 2x)}}{{{P_t}}}, \\
&{d_{FD_a}} = \frac{1}{{{P_t}}}\frac{{{C_3}\frac{x}{{1 + {\eta _1}x}}{e^{ - {C_1}x}}}}{{1 - \frac{1}{{1 + {\eta _1}x}}{e^{ - {C_1}x}}}} \approx \frac{{1 - \frac{x}{{{P_t}}}{C_3}}}{{1 + \frac{{{P_t}{\eta _1}}}{{{C_3}}}}} \le {1 - \frac{x}{{{P_t}}}{C_3}} ,\\
&{d_{HY}} \approx 1 - \frac{1}{{{P_t}}}\frac{{ {{{({C_3}x)}^2} + {\eta _1}x{{\left( {{C_3}({x^2} + 2x) + \frac{{x + 1}}{{{\eta _1}{\lambda _1}}}} \right)}^2}} }}{{{C_3}x + {\eta _1}x{C_3}({x^2} + 2x) + {\eta _1}x\frac{{x + 1}}{{{\eta _1}{\lambda _1}}}}}.
\end{align}\end{small}
\vspace{-5mm}

\begin{remark}\label{remark:1}
When SNR goes infinite, the diversity order of HD and HY scheme approaches one.
Thus, the X-duplex relay achieves nearly double diversity order compared with HD mode and HY scheme at high SNR.
\end{remark}

\vspace{-2mm}
\section{Simulation Results}

In this section, we present the performance of the X-duplex relay system.
Without loss of generality, we assume equal power allocation $P_S =P_R$, set all channel gains $\lambda_i$ to one, and $\eta =\eta_1 =\eta_2$.
We consider BPSK modulation and set the threshold $R_0$ as 2 bps/Hz.

%Fig.~\ref{fig:outage} and Fig.~\ref{fig:ser} plots the numerical and analytical results of the outage probability and SER performance of the X-duplex relay system respectively.
%The performance of pure FD or HD mode, the joint relay and antenna mode selection scheme (RAMS) proposed in \cite{Ky}, and the HY scheme proposed in \cite{Ik} are plotted for comparison.
%The simulated outage probability and SER curves tightly match with the expression in \eqref{eq:cdf} and \eqref{eq:ser}, respectively.
%It can be observed that the proposed X-duplex considerably improves system performance and outperforms the other schemes.
%In the medium SNR, the diversity order of X-duplex is higher than the pure FD or HD mode, and HY scheme.
%At high SNR, the performance floor in FD mode and RAMS scheme caused by RSI is removed in the X-duplex scheme.

\begin{figure}
\centering
\includegraphics[width=5in]{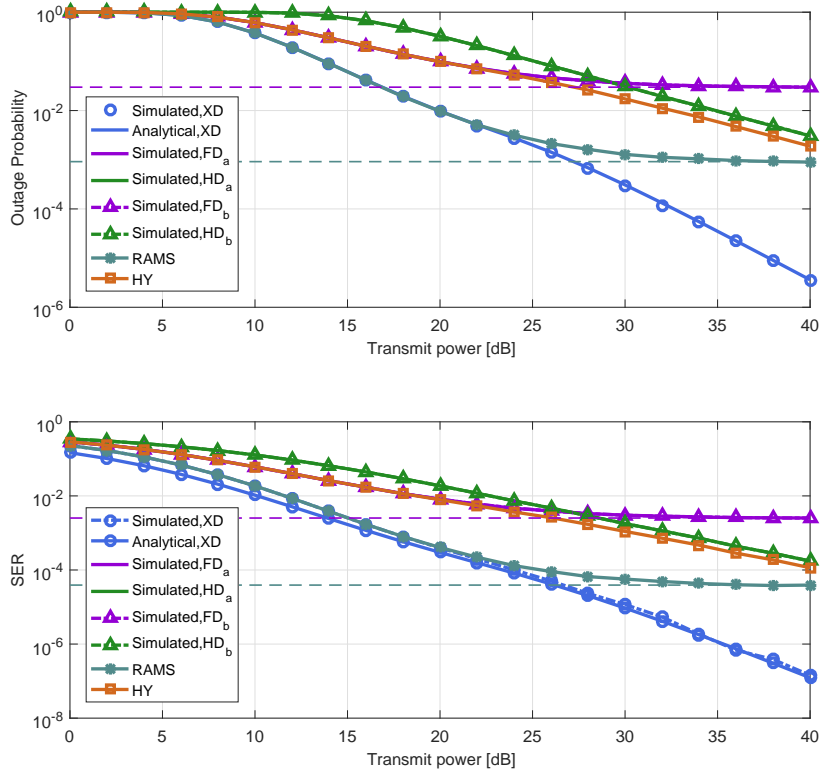}%SER.eps
%\vspace{-3mm}
\caption{\label{fig:ser} Outage probability and average SER of X-duplex relay system versus the transmit power when $\eta {\rm{ = 0}}{\rm{.01}}$.}
%\vspace{-4mm}
\end{figure}

Fig.~\ref{fig:ser} plots the numerical and analytical results of the outage probability and average SER performance of the X-duplex relay system.
The performance of pure FD or HD mode, HY \cite{Trhybird}, and RAMS \cite{Ky} are plotted for comparison.
The simulated outage probability and SER curves tightly match with the expressions in \eqref{pout}, \eqref{eq:ser}.
It can be observed that the proposed X-duplex considerably improves system performance and outperforms the other schemes.
In the medium SNR, the diversity order of X-duplex is higher than the pure FD or HD mode, and HY scheme.
At high SNR, the performance floor in FD mode and RAMS scheme caused by RSI is significantly reduced in the X-duplex scheme.
This is because the X-duplex benefits from the HD mode, whose performance is irrelevant to RSI and improves with the increase of transmit power, thus the impact of the performance floor in FD mode on X-duplex relaying system is mitigated with the increase of SNR.
%At high SNR, the performance floor in FD mode and RAMS scheme caused by RSI is significantly reduced in the X-duplex scheme.
%This is because the X-duplex takes the benefits of two HD modes, which are more likely to be selected at high SNR, thus the influence of performance floor in FD mode is mitigated.

%Fig.~\ref{fig:ser} plots the numerical and analytical results of the average SER performance of the X-duplex relay system.
%The performance of pure FD or HD mode, the joint relay and antenna mode selection scheme (RAMS) proposed in \cite{Ky}, and the HY scheme proposed in \cite{Trhybird} are plotted for comparison.
%The simulated SER curve tightly matches with the expression in \eqref{eq:ser}.
%It can be observed that the proposed X-duplex considerably improves system performance and outperforms the other schemes.
%In the medium SNR, the diversity order of X-duplex is higher than the pure FD or HD mode, and HY scheme.
%At high SNR, the performance floor in FD mode and RAMS scheme caused by RSI is significantly reduced in the X-duplex scheme.

\begin{figure}
\centering
\includegraphics[width=5in]{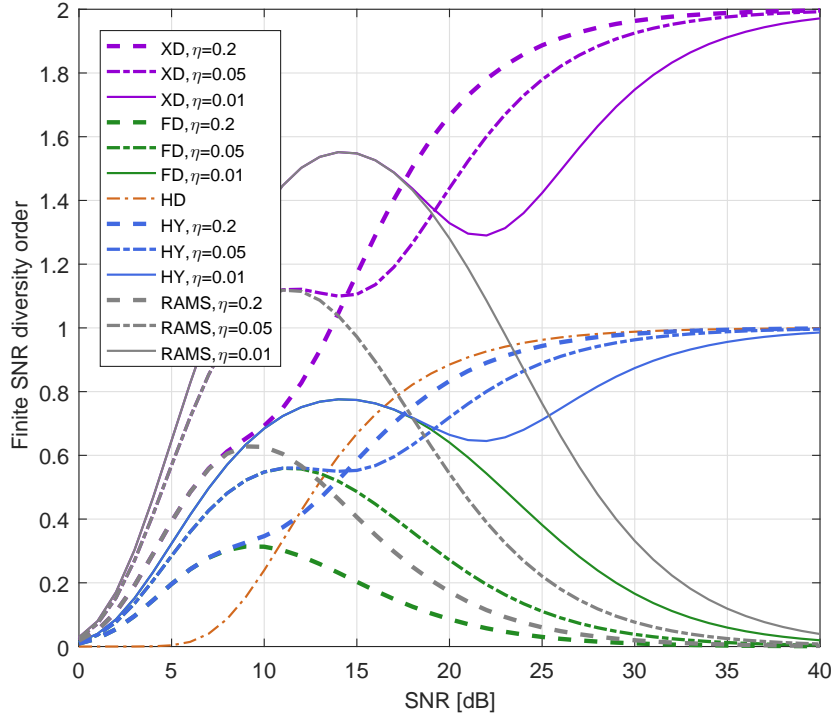}
%\vspace{-3mm}
\caption{\label{fig:diversityorder} Finite SNR diversity order versus the link SNR.}
%\vspace{-4mm}
\end{figure}

Fig.~\ref{fig:diversityorder} compares the finite SNR diversity order of the X-duplex scheme with the conventional pure FD and HD mode, RAMS and HY scheme.
The diversity order of X-duplex scheme is higher than other schemes and approaches two at high SNR, which is twice that of HD mode or HY scheme with fixed antennas, which is consistent with remark \ref{remark:1}.
We can observe that the diversity order of FD mode and RAMS scheme increases to the extreme point in medium SNR, where the influence of RSI on the SINR of FD mode is still small and limited.
The curve of RAMS approaches that of X-duplex as FD is more likely to be selected in this region.
As SNR continually increases, the impact of RSI on the FD mode gets more severe and the diversity order of FD and RAMS gradually decreases.
It is shown that the diversity order of FD mode and RAMS decreases to zero at high SNR due to RSI, thus the performance floor exists at high SNR.
By adaptively switching among the four modes, the X-duplex scheme significantly reduces the performance floor and achieves additional spatial diversity.

%\vspace{-2mm}
%\subsection{Performance and Complexity}
%In the X-duplex relaying system, the CSIs of five channels needs to be estimated and transmitted to the decision node through feedback channels, while there are only three channels in the HY scheme.
%The proposed X-duplex requires $\left\lceil {{{\log }_2}(5) - {{\log }_2}(3)} \right\rceil  = 1$ more bit in the feedback channel compared with HY.
%On the other side, the proposed X-duplex relay achieves nearly double diversity order compared with HD mode and HY scheme at high SNR.
%\begin{figure}
%\centering
%\includegraphics[width=3.3in]{outage.eps}
%\vspace{-1mm}
%\caption{\label{fig:outage} Outage probability of X-duplex relay system when $\eta {\rm{ = 0}}{\rm{.01}}$}
%\vspace{-4mm}
%\end{figure}

%\vspace{-2mm}
\section{Conclusions}
In this letter, we proposed a joint transmission mode and Tx/Rx antenna configuration scheme for the relay network where the relay is equipped with two antennas capable of transmission or reception.
In the proposed scheme, the relay adaptively configures its Tx/Rx antenna and duplex mode to minimize the SER.
The asymptotic average SER expression and diversity order were derived and validated by simulations.
Both analysis and simulations demonstrated that the X-duplex scheme improves the system performance, achieves almost twice diversity order and significantly reduces the performance floor compared to conventional relay schemes.

\section*{Appendix A: Proof of Proposition 1}
First, the set $\{ {{\gamma _{\max }} < x} \}$ can be transformed into $\{ {{\gamma _{f{d_a}}} < x,{\gamma _{f{d_b}}} < x,{\gamma _{h{d_a}}} < {x^2} + 2x,{\gamma _{h{d_b}}} < {x^2} + 2x} \}$.
As the probability of set $\{ {\gamma _{f{d_a}}} < x,{\gamma _{h{d_a}}} < {x^2} + 2x\}$ only contains the probabilities of ${\gamma _1}, {\gamma _4}, {\gamma ^1}_{SI}$, which are independent from ${\gamma _2}, {\gamma _3}, {\gamma ^2}_{SI}$ in $\{ {\gamma _{f{d_b}}} < x,{\gamma _{h{d_b}}} < {x^2} + 2x\}$.
The probability ${P^ * } = \Pr({\gamma _{\max }} < x) $ can be further written as
\begin{small}\begin{equation}\label{eq:px}
\hspace{-0.5mm}P^ * \hspace{-1mm}=\hspace{-1mm} \Pr({\gamma _{f{d_a}}}\hspace{-1mm} < x,{\gamma _{h{d_a}}}\hspace{-1mm} < {x^2} + 2x) \cdot\hspace{-0.5mm} \Pr({\gamma _{f{d_b}}}\hspace{-1mm} < x,{\gamma _{h{d_b}}}\hspace{-1mm} < {x^2} + 2x).
\end{equation}
\end{small}
\vspace{-1mm}
Denoting $P^ {1}=\Pr({\gamma _{f{d_a}}} < x,{\gamma _{h{d_a}}} < {x^2} + 2x)$, we can write
\begin{small}\begin{align}
&\Pr({\gamma _{f{d_a}}} < x,{\gamma _{h{d_a}}} < {x^2} + 2x) = 1 - \Pr({\gamma _{f{d_a}}} > x) \nonumber\\
&\hspace{3mm} - \Pr({\gamma _{h{d_a}}} > {x^2} + 2x) + \Pr({\gamma _{f{d_a}}} > x,{\gamma _{h{d_a}}} > {x^2} + 2x).
\end{align}\end{small}

The probability \begin{small}$\Pr({\gamma _{f{d_a}}} > x) = \Pr (({X_1} - x)({P_R}{\gamma _4} - x) > {x^2} + x)$ \end{small}can be derived as
\vspace{-1mm}
\begin{small}\begin{equation}
\Pr({\gamma _{f{d_a}}} > x) = \frac{1}{{{\lambda _4}}}\int\limits_{x/{P_R}}^\infty  {\frac{{{e^{ - \frac{1}{{{P_S}{\lambda _1}}}(x + \frac{{{x^2} + x}}{{{P_R}{\gamma _4} - x}}) - \frac{1}{{{\lambda _4}}}{\gamma _4}}}}}{{1 + \eta_1 (x + \frac{{{x^2} + x}}{{{P_R}{\gamma _4} - x}})}}} d{\gamma _4},
\end{equation}\end{small}
\vspace{-2mm}
at high SNR, we use the following approximation
\begin{small}\begin{equation}\label{eq:appro}
\frac{1}{{1 + \eta_1 (x + \frac{{{x^2} + x}}{{{P_R}{\gamma _4} - x}})}}\approx \frac{1}{{1 + \eta_1 x}}(1 - \frac{\eta_1 }{{1 + \eta_1 x}}\frac{{{x^2} + x}}{{{P_R}{\gamma _4} - x}}),
\end{equation}\end{small}
%With approximation \begin{small}$\frac{1}{{1 + \eta_1 (x + \frac{{{x^2} + x}}{{{P_R}{\gamma _4} - x}})}}\approx \frac{1}{{1 + \eta_1 x}}(1 - \frac{\eta_1 }{{1 + \eta_1 x}}\frac{{{x^2} + x}}{{{P_R}{\gamma _4} - x}})$\end{small} at high SNR and
With \cite[eq.(3.471.9)]{zwillinger2014table}, $\Pr({\gamma _{f{d_a}}} > x)$ is obtained.
The probabilities $\Pr({\gamma _{h{d_a}}} > x^2 + 2x)$  and $\Pr({\gamma _{f{d_a}}} > x)$ can be derived as
\begin{small}\begin{align}\label{eq:hda}
&\Pr ({\gamma _{h{d_a}}} > {x^2} + 2x){\rm{ = }}\beta _0^1{K_1}(\beta _0^1){e^{ - C_1({x^2} + 2x)}},\\\label{eq:fda}
&\Pr ({\gamma _{f{d_a}}} > x) \approx \frac{{\beta _1^1{K_1}(\beta _1^1)}}{{1 + {\eta _1}x}}{e^{ - {C_1}x}} - {\alpha ^1}{K_0}(\beta _1^1){e^{ - {C_1}x}},
 \end{align}\end{small}
where \begin{small}$\beta _0^1 = 2\sqrt {\frac{{{{({x^2} + 2x)}^2} + {x^2} + 2x}}{{{\lambda _1}{\lambda _4}{P_S}{P_R}}}}$, ${\alpha ^1} = \frac{{2{\eta _1}({x^2} + x)}}{{{\lambda _4}{P_R}{{(1 + {\eta _1}x)}^2}}}$.\end{small}

The set $\{ {\gamma _{f{d_a}}} > x,{\gamma _{h{d_a}}} > {x^2} + 2x\} $ can be transformed into $\{ {\gamma _4} > \frac{x}{{{P_R}}},{\gamma _4} > \frac{{{x^2} + 2x}}{{{P_R}}},{\gamma _1} > \frac{{{P_R}\gamma _{SI}^1 + 1}}{{{P_S}}}h({\gamma _4}),{\gamma _1} > \frac{1}{{{P_S}}}g({\gamma _4})\} $ where $h(\gamma ) = x + \frac{{x + {x^2}}}{{{P_R}\gamma  - x}}$ , $g(\gamma ) = {x^2} + 2x + \frac{{{{({x^2} + 2x)}^2} + {x^2} + 2x}}{{{P_R}\gamma  - {x^2} - 2x}}$.
As the value of $\gamma_{fd_a}, {\gamma _{hd_a}}$ are positive definite, we only consider the case when $x > 0$.
Therefore, the set $\{ {\gamma _{fd_a}} > x,{\gamma _{hd_a}} > {x^2 + 2x}\} $ can be further simplified as $\{ {\gamma _4} > \frac{{{x^2} + 2x}}{{{P_R}}},{\gamma _1} > \frac{{{P_R}\gamma _{SI}^1 + 1}}{{{P_S}}}h({\gamma _4}),{\gamma _1} > \frac{1}{{{P_S}}}g({\gamma _4})\} $.

We define \begin{small}$\delta = \frac{1}{{{P_S}}}\left[ {({P_R}\gamma _{SI}^1 + 1)h({\gamma _2}) - g({\gamma _2})} \right]$\end{small}.
%\begin{small}\begin{equation}
%\delta = \frac{1}{{{P_S}}}\left[ {({P_R}\gamma _{SI}^1 + 1)h({\gamma _2}) - g({\gamma _2})} \right],
%\end{equation}\end{small}
%\hspace{-1mm}
when $\delta  > 0$ , ${\gamma _{SI}^1} > \frac{{x{\gamma _4} + {\gamma _4}}}{{{P_R}{\gamma _4} - {x^2} - 2x}}$ ,when $\delta < 0$ , $0< {\gamma _{SI}^1} < \frac{{x{\gamma _4} + {\gamma _4}}}{{{P_R}{\gamma _4} - {x^2} - 2x}}$.
Thus, \begin{small}$\Pr \{ {\gamma _{fd_a}} > x,{\gamma _{hd_a}} > {x^2 + 2x}\}$ splits into two sub-probabilities, ${L_1} = \Pr \{ {\gamma _4} > \frac{{{x^2} + 2x}}{{{P_R}}},{\gamma _1} > \frac{{{P_R}{\gamma _{SI}^1} + 1}}{{{P_S}}}h({\gamma _4}),{\gamma _{SI}^1} > \frac{{x{\gamma _4} + {\gamma _4}}}{{{P_R}{\gamma _4} - {x^2} - 2x}}\}$\end{small} and \begin{small}${L_2} = \Pr \{ {\gamma _4} > \frac{{{x^2} + 2x}}{{{P_R}}},{\gamma _1} > \frac{1}{{{P_S}}}g({\gamma _4}),0 < {\gamma _{SI}^1} < \frac{{x{\gamma _4} + {\gamma _4}}}{{{P_R}{\gamma _4} - {x^2} - 2x}}\}$\end{small}.
%Denoting \begin{small}${L_1} = \int\limits_{\frac{{{x^2} + 2x}}{{{P_R}}}}^\infty  {{f_{{\gamma _4}}}\int\limits_{\frac{{x{\gamma _4} + {\gamma _4}}}{{{P_R}{\gamma _4} - {x^2} - 2x}}}^\infty  {{f_{\gamma _{SI}^1}}} } \int\limits_{\frac{{{P_R}\gamma _{SI}^1 + 1}}{{{P_S}}}{\rm{h}}({\gamma _4})}^\infty  {{f_{{\gamma _1}}}d{\gamma _1}d\gamma _{SI}^1} d{\gamma _4}$\end{small},
%\begin{small}\begin{equation}\label{1}
%  {L_1} = \int\limits_{\frac{{{x^2} + 2x}}{{{P_R}}}}^\infty  {{f_{{\gamma _4}}}\int\limits_{\frac{{x{\gamma _4} + {\gamma _4}}}{{{P_R}{\gamma _4} - {x^2} - 2x}}}^\infty  {{f_{\gamma _{SI}^1}}} } \int\limits_{\frac{{{P_R}\gamma _{SI}^1 + 1}}{{{P_S}}}{\rm{h}}({\gamma _4})}^\infty  {{f_{{\gamma _1}}}d{\gamma _1}d\gamma _{SI}^1} d{\gamma _4},
%\end{equation}
%\end{small}
%\hspace{-1mm}
%where ${{f_{{\gamma _1}}}}$, ${{f_{{\gamma _4}}}}$, ${{f_{{\gamma _{SI}^1}}}}$ are the probability distribution function (PDF) of $\gamma_1$, $\gamma_4$, ${\gamma _{SI}^1}$.
With the approximation in \eqref{eq:appro}, and $\frac{1}{{{\gamma _4} - x/{P_R}}} \approx \frac{1}{{{\gamma _4} - ({x^2 + 2x})/{P_R}}}$ at high SNR, \cite[eq.(3.324.1)]{zwillinger2014table} and \cite[eq.(3.462.20)]{zwillinger2014table}, ${L_1}$, $L_2$ can be derived as
\setlength{\arraycolsep}{0.0em}
\begin{small}\begin{eqnarray}\label{eq:L1}
{L_1} &&= \frac{{\beta _2^1}}{{1 + {\eta _1}x}}{K_1}({\beta _2^1}){e^{ - \beta _3^1}} - {\alpha ^1} {K_0}(\beta _2^1){e^{ - \beta _3^1}},\\\label{eq:L2}
{L_2} &&= \beta _0^1{K_1}(\beta _0^1){e^{ - {C_1}({x^2} + 2x)}} - \beta _2^1{K_1}(\beta _2^1){e^{ - \beta _3^1}}.
\end{eqnarray}\end{small}

With \eqref{eq:hda}, \eqref{eq:fda}, \eqref{eq:L1}, \eqref{eq:L2}, the probability $P^ {1}$ is given as
\begin{small}\begin{align}\label{eq:hs}
{P^1} &= 1 - \frac{1}{{1 + {\eta _1}x}}\left[ {\beta _1^1{K_1}(\beta _1^1){e^{ - {C_1}x}} + {\eta _1}x\beta _2^1{K_1}(\beta _2^1){e^{ - \beta _3^1}}} \right]\nonumber\\
& + \frac{{2{\eta _1}({x^2} + x)}}{{{\lambda _4}{P_R}{{(1 + {\eta _1}x)}^2}}}\left[ {{K_0}(\beta _1^1){e^{ - {C_1}x}} - {K_0}(\beta _2^1){e^{ - \beta _3^1}}} \right].
\end{align}\end{small}
\hspace{0.5mm}Similarly $P^ {2}=\Pr({\gamma _{f{d_b}}} < x,{\gamma _{h{d_b}}} < {x^2} + 2x)$ can be derived.
With \eqref{eq:px}, proposition 1 is proved.
Due to the approximations used in \eqref{eq:fda}, \eqref{eq:L1}, the derived CDF \eqref{eq:cdf} is an approximate and asymptotic expression and is quite accurate at high SNR.

\section*{Appendix B: Proof of Proposition~2}
After substituting (\ref{eq:cdf}) into (\ref{ser}) and adopting the approximation in the high SNR region that ${K_1}(z)$ converges to $\frac{1}{z}$, and that the value of ${K_0}(z)$ is comparatively small \cite[eq.(10.30)]{Frank}, which can be ignored for asymptotic analysis.
We can derive
\setlength{\arraycolsep}{0.0em}
\begin{small}\begin{align}\label{serpart}
&\overline {SER} \approx \frac{{{a_1}\sqrt {{a_2}} }}{{2\sqrt \pi  }}\int\limits_0^\infty  {\frac{{{e^{ - {a_2}x}}}}{{\sqrt x }}\left\{ {1 - \frac{{{e^{ - {C_1}x}} + {\eta _1}x{e^{ - \beta _3^1}}}}{{1 + {\eta _1}x}} - \frac{{{e^{ - {C_2}x}} + {\eta _2}x{e^{ - \beta _3^2}}}}{{1 + {\eta _2}x}}} \right.}  \nonumber\\
 & \left. { + \frac{{{e^{ - {C_1}x - {C_2}x}} + {\eta _1}x{e^{ - \beta _3^1 - {C_2}x}} + {\eta _2}x{e^{ - {C_1}x - \beta _3^2}} + {\eta _1}{\eta _2}x^2{e^{ - \beta _3^1\beta _3^2}}}}{{(1 + {\eta _1}x)(1 + {\eta _2}x)}}} \right\}dx.
\end{align}\end{small}
\vspace{-3mm}

With \cite[eq.(3.381.4)]{zwillinger2014table}, $S_1 =\int\limits_0^\infty  {\frac{{{e^{ - {a_2}x}}}}{{\sqrt x }}dx = {a_2}^{ - \frac{1}{2}}\Gamma (\frac{1}{2})}$.

With \cite[eq.(3.383.10)]{zwillinger2014table}, $S_2 =\int\limits_0^\infty  {\frac{{{e^{ - {a_2}x - {C_1}x}}}}{{\sqrt x (1 + {\eta _1}x)}}dx} $ is given as
\vspace{-2mm}
\begin{small}\begin{equation}
\hspace{-1mm} S_2 =\hspace{-1mm}  \frac{1}{\eta_1 }\int\limits_0^\infty  {\frac{{{e^{ - ({a_2} + C_1)x}}}}{{\sqrt x (\frac{1}{\eta_1 } + x)}}dx}  =\hspace{-1mm}  \sqrt {\frac{\pi }{{{\eta _1}}}} {e^{\frac{1}{{{\eta _1}}}({a_2} + {C_1})}}\Gamma (\frac{1}{2},\frac{{{a_2} + {C_1}}}{{{\eta _1}}}).\vspace{-4mm}
\end{equation}\end{small}

Denoting \begin{small}$S_3 = \int\limits_0^\infty  {\frac{{{\eta _1}x \cdot {e^{ - {a_2}x - \beta _3^1}}}}{{\sqrt x (1 + {\eta _1}x)}}dx}$\end{small}, when the SNR is high and $x$ is around zero, approximation $\frac{1}{{1 + x}} \approx {e^{ - x}} + \frac{1}{2}{x^2}{e^{ - \frac{5}{3}x}}$ \cite{Ky} is used, with \cite[eq.(3.462.1)]{zwillinger2014table},  $S_3$ is given as
 \setlength{\arraycolsep}{0.0em}
\begin{small}\begin{eqnarray}\label{ser4}
 {S_3}  &&\approx  \int\limits_0^\infty  {{\eta _1}\sqrt x ({e^{ - {\eta _1}x}} + \frac{1}{2}{\eta _1}^2{x^2}{e^{ - \frac{5}{3}{\eta _1}x}}){e^{ - {a_2}x - \beta _3^1}}dx} \nonumber\\
 &&={\eta _1}{e^{ - \frac{1}{{{\lambda _1}{P_S}{\eta _1}}}{\rm{ + }}\frac{{{\mu _1}^2}}{{8C_1}}}}{\left( {2C_1} \right)^{ - \frac{3}{4}}}\Gamma (\frac{3}{2}){D_{ - \frac{3}{2}}}(\frac{{{\mu _1}}}{{\sqrt {2C_1} }}) \nonumber\\
&&+ \frac{1}{2}{\eta _1}^3{e^{ - \frac{1}{{{\lambda _1}{P_S}{\eta _1}}}{\rm{ + }}\frac{{{\mu _2}^2}}{{8C_1}}}}{\left( {2C_1} \right)^{ - \frac{7}{4}}}\Gamma (\frac{7}{2}){D_{ - \frac{7}{2}}}(\frac{{{\mu _2}}}{{\sqrt {2C_1} }}).
 \end{eqnarray}\end{small}

\vspace{-3mm}
Similarly, $S_4 = \int\limits_0^\infty  {\frac{{{e^{ - {a_2}x}}}}{{\sqrt x }}\frac{{{e^{ - {C_2}x}} + {\eta _2}x{e^{ - \beta _3^2}}}}{{1 + {\eta _2}x}}dx}$ can be derived.
For the last part denoted as $S_5$ in \eqref{serpart}, with some mathematical manipulations, the value can be also derived.
Substituting $S_1$, $S_2$, $S_3$, $S_4$, $S_5$  into \eqref{serpart}, \eqref{eq:ser} can be obtained.
Therefore, proposition 2 is proved.

\end{document}